\documentclass[aps,superscriptaddress]{revtex4-2}
\usepackage[colorlinks=true, pdfstartview=FitV, linkcolor=blue, citecolor=red, urlcolor=magenta, breaklinks=true]{hyperref}
\usepackage{graphicx}  
\usepackage{amsfonts}
\begin{document}
\title{Hawking radiation and stability of the canonical acoustic black holes}
\author{M. A. Anacleto}
\email{anacleto@df.ufcg.edu.br}
\affiliation{Departamento de F\'{\i}sica, Universidade Federal de Campina Grande
Caixa Postal 10071, 58429-900 Campina Grande, Para\'{\i}ba, Brazil}
\author{F. A. Brito}
\email{fabrito@df.ufcg.edu.br}
\affiliation{Departamento de F\'{\i}sica, Universidade Federal de Campina Grande
Caixa Postal 10071, 58429-900 Campina Grande, Para\'{\i}ba, Brazil}
\affiliation{Departamento de F\'isica, Universidade Federal da Para\'iba, 
Caixa Postal 5008, 58051-970 Jo\~ao Pessoa, Para\'iba, Brazil}
\author{E. Passos}
\email{passos@df.ufcg.edu.br}
\affiliation{Departamento de F\'{\i}sica, Universidade Federal de Campina Grande
Caixa Postal 10071, 58429-900 Campina Grande, Para\'{\i}ba, Brazil}

\begin{abstract} 

In this paper we determine the Hawking temperature and entropy of a modified canonical acoustic black hole. 
In our results {we obtain an entropy-area relation that has a logarithmic
correction term in leading order.} 
We also analyze the stability condition of the noncommutative canonical acoustic black hole 
and also with quantum corrections implemented by the generalized Heisenberg uncertainty principle.

\end{abstract}
\maketitle
\pretolerance10000

\section{Introduction}
Black holes are fascinating objects predicted by General Relativity and have attracted enormous interest in high-energy physics, astrophysics and astronomy.
In addition, in the physics of black holes some phenomena have been extensively analyzed 
such as {Hawking radiation}~\cite{Kerner:2006vu,Akhmedov:2006pg}, superradiance and quasinormal modes.
{The thermodynamics of black holes with quantum corrections implemented by the generalized uncertainty principle 
was analyzed in~\cite{Ovgun:2015jna,Ovgun:2015box,Ovgun:2017hje,Javed:2018msn} 
and Hawking radiation of black holes in arbitrary dimensions with or without Gauss-Bonnet term have been investigated in~\cite{Kuang:2017sqa,Gonzalez:2017zdz,Ovgun:2019ygw}.}
It is well known that when a black hole interacts with the surrounding matter the disturbances cause the black hole to suffer oscillation, leading to the emission of gravitational waves~\cite{Konoplya:2011qq,Konoplya:2002zu,Maggiore:1999vm,Hughes:2014yia}. 
Such gravitational waves that are ripples in the spacetime fabric were detected on September 14, 2015 by the collaboration LIGO-VIRGO~\cite{LIGOScientific:2016aoc,LIGOScientific:2017vwq}.
Also, in 2019 the image of a supermassive black hole at the center of galaxy M87 was captured by the 
Event Horizon Telescope~\cite{event2019firstI,event2019firstVI}. 
Therefore, these observations usher in a new era in the development of black hole physics.

In order to examine Hawking radiation and other phenomena to understand quantum gravity, acoustic black holes were proposed for this purpose by Unruh in 1981~\cite{Unruh:1980cg,Unruh:1994je}. 
Since then, great advances have been achieved in the investigation of Hawking radiation in the theoretical field as well as in the experimental field.
As we know, an acoustic black hole can be built when the fluid motion becomes greater than the local speed of sound. 
Therefore, acoustic event horizons are formed by showing similar characteristics to the gravitational black hole event horizon. 
Moreover, these acoustic black holes, as well as analogous models, are quite useful in various branches of physics with applications, for example, in high energy physics, condensed matter and quantum physics~\cite{Visser:1997ux,Barcelo:2005fc}.
Experimentally, the Hawking temperature of acoustic black holes has been successfully measured~\cite{MunozdeNova:2018fxv,Isoard:2019buh}. This has also been done in other physical systems~\cite{Steinhauer:2014dra,Drori:2018ivu,Rosenberg:2020jde,Guo:2019tmr,Bera:2020doh,Blencowe:2020ygo}.
The first experimental study of Hawking radiation from acoustic black holes was carried out in the Bose-Einstein condensate system~\cite{Lahav:2009wx}.

In recent years, by considering Gross-Pitaevskii relativistic and Yang-Mills theories, effective acoustic black hole metrics for analogous gravity models have been constructed~\cite{Ge:2019our}. 
However, based on a holographic approach, the acoustic black hole of a D3-black brane was obtained~\cite{Yu:2017bnu}.
Furthermore, relativistic metrics of acoustic black holes have been obtained from the Abelian-Higgs model~\cite{Ge:2010wx,Anacleto:2010cr,Anacleto:2011bv,Anacleto:2013esa} and also from other models~\cite{Bilic:1999sq,Fagnocchi:2010sn,Giacomelli:2017eze,Visser:2010xv}.
Besides, by applying these relativistic effective metrics, many works have been carried out to explore, for example, 
{analogous Hawking radiation}~\cite{Sakalli:2016mnk}, superradiance~\cite{Basak:2002aw,Richartz:2009mi,Anacleto:2011tr,Zhang:2011zzh,Ge:2010eu}, entropy~\cite{Zhao:2012zz,Anacleto:2014apa,Anacleto:2015awa,Anacleto:2016qll,Anacleto:2019rfn}, quasinormal modes~\cite{Cardoso:2004fi,Nakano:2004ha,Berti:2004ju,Chen:2006zy,Guo:2020blq,Ling:2021vgk}, and as well as, in other models~\cite{Dolan:2011zza,Anacleto:2012ba,Anacleto:2012du,Anacleto:2015mta,Anacleto:2016ukc,Anacleto:2018acl,Anacleto:2020kxj,Anacleto:2021wmv,Qiao:2021trw,Ribeiro:2021fpk}. 
In addition, analogous Hawking radiation and the analytical solution of a massless scalar field have been addressed in~\cite{Vieira:2014rva}. 
Although, due to equivalent mathematical descriptions between acoustic black holes and gravitational black holes, it has been conjectured that if a physical phenomenon occurs in gravitational black holes, it can also happen in acoustic black holes. 
In this sense, in~\cite{Zhang:2016pqx}, it has been shown that there is a thermodynamic-like description for acoustic black holes in two dimensions; entropy and specific heat capacity have also been calculated.
Hence, unlike Hawking radiation, an analogous form for the Bekenstein-Hawking entropy has been little investigated so far. 
However, it was explored in~\cite{Rinaldi:2011nb}, that such an analogy arises in a Bose-Einstein condensate system. Thus, the Bekenstein-Hawking entropy can be understood as an entanglement entropy. 
This entanglement entropy is related to the phonons generated in the Hawking mechanism. 
Furthermore, the dependence of entropy on the area of the event horizon of the acoustic black hole was examined in~\cite{Steinhauer:2015ava}. 
Moreover, the entanglement entropy of an acoustic black hole was also explored in 1D degenerate ideal fermi fluids~\cite{Giovanazzi:2011az}. 
So, with the examples mentioned above in ultracold quantum gases, we can look at other systems that exhibit entanglement entropy.

Therefore, guided by all these works, in this paper, we are interested in examining whether such a conjecture occurs in models of canonical acoustic black holes with modified metrics. 
In this sense, we will address the effect of noncommutativity on the computation of Hawking radiation, entropy, and stability of the canonical acoustic black hole, as well as the effect of minimum length on such quantities. 
As a result, we show that for a given minimum radius (which depends on the noncommutative parameter) the noncommutative canonical acoustic black hole becomes stable. In the absence of noncommutativity the acoustic black hole presents instability. 
In addition, we verify that for a certain minimum radius that depends on the deformation parameter, the canonical acoustic black hole with minimum length effect also becomes stable.

The paper is organized as follows. In Sec.~\ref{II} we  briefly review the steps to find the noncommutative acoustic black hole metrics and we calculate the Hawking temperature, entropy and specific heat capacity of the noncommutative canonical acoustic black hole. 
In Sec.~\ref{qcbh} we compute the Hawking temperature, entropy and specific heat capacity of the 
canonical acoustic black hole with quantum correction.
Finally in Sec.~\ref{conclu} we present our final considerations.

\section{noncommutative canonical acoustic black hole}
\label{II}
The metric of a noncommutative canonical acoustic black hole has been found by us in~\cite{Anacleto:2011bv}.
Here we will briefly show the steps to obtain the noncommutative acoustic black hole metric in (3+1) dimensions from quantum field theory. 
We start from the the Lagrangian of the noncommutative Abelian Higgs model in flat space given by~\cite{Ghosh:2004wi}
\begin{eqnarray}
\label{acao}
\hat{\cal L}&=&-\frac{1}{4}F_{\mu\nu}F^{\mu\nu}\left(1+\frac{1}{2}\theta^{\alpha\beta}F_{\alpha\beta}\right) 
+\left(1-\frac{1}{4}\theta^{\alpha\beta}F_{\alpha\beta}\right)\left(|D_{\mu}\phi|^2+ m^2|\phi|^2-b|\phi|^4\right)
\nonumber\\
&+&\frac{1}{2}\theta^{\alpha\beta}F_{\alpha\mu}\left[(D_{\beta}\phi)^{\dagger}D^{\mu}\phi+(D^{\mu}\phi)^{\dagger}D_{\beta}\phi \right],
\end{eqnarray}
where $F_{\mu\nu}=\partial_{\mu}A_{\nu}-\partial_{\nu}A_{\mu}$,  $D_{\mu}\phi=\partial_{\mu}\phi - ieA_{\mu}\phi$  
and the parameter $\theta^{\alpha\beta}$ is a constant, real-valued antisymmetric $D\times D$- matrix in $D$-dimensional spacetime with dimensions of length squared.

Now, we applying  $\phi = \sqrt{\rho(x, t)} \exp {(iS(x, t))}$ into the original Lagrangian, we have
\begin{eqnarray}
\label{lagran}
{\cal L}&=&-\frac{1}{4}F_{\mu\nu}F^{\mu\nu}\left(1-2\vec{\theta}\cdot\vec{B}\right)
+\rho(\tilde{\theta}g^{\mu\nu}+\Theta^{\mu\nu}){\cal D}_{\mu}S{\cal D}_{\nu}S+\tilde{\theta} m^2\rho-\tilde{\theta}b\rho^2
+\frac{\rho}{\sqrt{\rho}}(\tilde{\theta}g^{\mu\nu}+\Theta^{\mu\nu})\partial_{\mu}\partial_{\nu}\sqrt{\rho},
\end{eqnarray}
where ${\cal D}_{\mu}=\partial_{\mu}-eA_{\mu}/S  $, $\tilde{\theta}=(1+\vec{\theta}\cdot\vec{B})$, $\vec{B}=\nabla\times\vec{A}$ and $\Theta^{\mu\nu}=\theta^{\alpha\mu}{F_{\alpha}}^{\nu}$. 
In our analysis we consider the case where there is no noncommutativity between space and time, that is $\theta^{0i}=0$ and use $\theta^{ij}=\varepsilon^{ijk}\theta^{k}$, $F^{i0}=E^{i}$ and $F^{ij}=\varepsilon^{ijk}B^{k}$.

Hence, by linearizing the equations of motion around the background $(\rho_0,S_0)$, with $\rho=\rho_0+\rho_1$ and  $S=S_0+\psi$ we find the equation of motion for a linear acoustic disturbance $\psi$ given by 
\begin{eqnarray}
\frac{1}{\sqrt{-g}}\partial_{\mu}(\sqrt{-g}g^{\mu\nu}\partial_{\nu})\psi=0,
\end{eqnarray}
where $g_{\mu\nu}=\frac{b\rho_0}{2c_s\sqrt{f}}\tilde{g}_{\mu\nu}$ is the relativistic acoustic metric with noncommutative corrections in (3+1) dimensions and with $ \tilde{g}_{\mu\nu} $ given in the form~\cite{Anacleto:2011bv}
\begin{eqnarray}
\tilde{g}_{tt}&=&-[(1-3\vec{\theta}\cdot\vec{B})c^{2}_{s}-(1+3\vec{\theta}\cdot\vec{B})v^2
+2(\vec{\theta}\cdot\vec{v})(\vec{B}\cdot\vec{v})-(\vec{\theta}\times\vec{E})\cdot\vec{v}],
\nonumber\\
\tilde{g}_{tj}&=&-\frac{1}{2}(\vec{\theta}\times\vec{E})^{j}(c^{2}_{s}+1)-\left[2(1+2\vec{\theta}\cdot\vec{B})
-(\vec{\theta}\times\vec{E})\cdot\vec{v}\right]\frac{v^j}{2}+\frac{B^j}{2}(\vec{\theta}\cdot\vec{v})+\frac{\theta^j}{2}(\vec{B}\cdot\vec{v}),
\nonumber\\
\tilde{g}_{it}&=&-\frac{1}{2}(\vec{\theta}\times\vec{E})^{i}(c^{2}_{s}+1)-\left[2(1+2\vec{\theta}\cdot\vec{B})-(\vec{\theta}\times\vec{E})\cdot\vec{v}\right]\frac{v^i}{2}
+\frac{B^i}{2}(\vec{\theta}\cdot\vec{v})+\frac{\theta^i}{2}(\vec{B}\cdot\vec{v}),
\nonumber\\
\tilde{g}_{ij}&=&[(1+\vec{\theta}\cdot\vec{B})(1+c^2_{s})-(1+\vec{\theta}\cdot\vec{B})v^2
-(\vec{\theta}\times\vec{E})\cdot\vec{v}]\delta^{ij}+(1+\vec{\theta}\cdot\vec{B})v^{i}v^{j}.
\nonumber\\
f&=&[(1-2\vec{\theta}\cdot\vec{B})(1+c^2_{s})-(1+4\vec{\theta}\cdot\vec{B})v^2]
-3(\vec{\theta}\times\vec{E})\cdot\vec{v}+2(\vec{B}\cdot\vec{v})(\vec{\theta}\cdot\vec{v}).
\end{eqnarray}
Here, in deducing the metric, we assume linear perturbations in the scalar sector and keep the vector field $A_{\mu}$ unchanged.

In the following we shall focus on the noncommutative canonical acoustic black hole metrics in (3+1) dimensions~\cite{Anacleto:2011bv} to address the issues of Hawking temperature and entropy. 
Now we consider an incompressible fluid with spherical symmetry. In this case the density $\rho$ is a position independent quantity and the continuity equation implies that $v\sim\frac{1}{r^2}$. The sound speed is also a constant.

The noncommutative acoustic metric can be written as a Schwarzschild metric type, up to an irrelevant
position-independent factor, in the nonrelativistic limit as follows,
\begin{eqnarray}
ds^2&=&-\tilde{{\cal F}}(v_{r})d\tau^2+\frac{[v_{r}^2\Gamma+\Sigma+\tilde{{\cal F}}(v_{r})\Lambda]}{\tilde{\cal F}(v_{r})}dr^2
+\frac{r^2(d{\vartheta}^2+\sin^2\vartheta d\phi^2)}{\sqrt{f}},
\end{eqnarray}
where
\begin{eqnarray}
\tilde{{\cal F}}(v_{r})&=&\frac{{\cal F}(v_{r})}{\sqrt{f(v_r)}}=\frac{1}{\sqrt{f(v_r)}}
\left[(1-3\vec{\theta}\cdot\vec{B})c^2_{s}-(1+3\vec{\theta}\cdot\vec{B})v^2_{r}-\theta {\cal E}_r{v}_{r}
+2(\theta_{r}B_{r}v^2_{r})\right],
\label{Fra}
\\
f(v_r)&=& 1-2\vec{\theta}\cdot\vec{B}-3\theta {\cal E}_r v_r,
\\
\Lambda(v_r)&=& 1+\vec{\theta}\cdot\vec{B}-\theta {\cal E}_r v_r,
\\
\Gamma(v_r)&=&1+4\vec{\theta}\cdot\vec{B}-2\theta {\cal E}_r {v}_r,
\\
\Sigma(v_r)&=&\left[\theta {\cal E}_r -({B}_r{v}_r)\theta_{r}-({\theta}_r {v}_r)B_{r}\right]v_{r},
\end{eqnarray}
being $ \theta {\cal E}_r=\theta(\vec{n}\times\vec{E})_{r} $. 
Now, by applying the relation $v_{r}=c_{s}\frac{r^2_{h}}{r^2}$, where $ r_h $ is the radius of the event horizon 
of the canonical acoustic black hole and making $ c_s=1 $ and so, the metric function of the noncommutative canonical acoustic black hole becomes
\begin{eqnarray}
\label{Fr}
\tilde{\cal F}(r)=\left[ 1-3\vec{\theta}\cdot\vec{B} -(1+3\vec{\theta}\cdot\vec{B}-2\theta_{r}B_{r})
\frac{r^4_{h}}{r^4}
-\theta {\cal E}_r\frac{r^2_{h}}{r^2}\right]
\left[1-2\vec{\theta}\cdot\vec{B}
-3\theta {\cal E}_r\frac{r^2_{h}}{r^2}\right]^{-1/2}.
\end{eqnarray}
Next, we will do our analysis considering the pure magnetic sector first and then we will investigate the pure electric sector.

\subsection{Magnetic Sector}

Hence, for $\theta_r=0$, $\vec{\theta}\cdot\vec{B}=\theta_3B_3\neq 0$,  $\theta {\cal E}_r=0$ (or  $E=0$)  with small $\theta_3B_3$,
\begin{equation} 
\label{Ttb}
T_{H}=\frac{\tilde{\cal F}^{\prime}(r_h)}{4\pi}=\frac{(1+3\theta_3B_3)}
{\sqrt{1-2\theta_3B_3}}\frac{1}{(\pi r_{h})}
=\frac{\left(1+4\theta_3B_3\right)}{(\pi r_{h})}.
\end{equation}
For $\theta=0$ the usual result is obtained. 
Here the temperature has its value increased when we vary the parameter $\theta$.

However, for the temperature in (\ref{Ttb}) we can find the entropy (entanglement
entropy~\cite{Anacleto:2019rfn}) of the modified canonical acoustic black hole by applying the first law of thermodynamics such that
\begin{eqnarray}
S=\int \frac{dE}{T}=\int \frac{dA}{4\pi r_h T_H}=\frac{(1-4\theta_3 B_3)}{4}A,
\end{eqnarray}
where $ A=4\pi r^2_h $ is the horizon area of the canonical acoustic black hole.

Now for  $\theta_r=0$, $\theta {\cal E}_r=0$ (or  $E=0$) and $\vec{\theta}\cdot\vec{B}=\theta_3B_3={\theta} r^2_0/r^2$ ($ r_0 $ is a parameter with length dimension and $ \theta $ is now dimensionless),   in (\ref{Fr}), we have
\begin{eqnarray}
\label{Fr2}
\tilde{\cal F}(r)=\left[ 1-3\theta\frac{r^2_0}{r^2} -\left(1+3\theta\frac{r^2_0}{r^2}\right)
\frac{r^4_{h}}{r^4} \right]
\left[1-2\theta\frac{r^2_0}{r^2}\right]^{-1/2}.
\end{eqnarray}
{At this point we call attention to the correction term $3\theta r^2_0/r^2$. 
This is a Reissner-Nordstr\"om type term with an effective charge $Q^2=3\theta r^2_0$ generated due to the noncommutativity.}

{For the Hawking temperature, we obtain}
\begin{eqnarray}
\label{THr0}
T_H=\frac{1}{\pi\left(r_h + 3\sqrt{\theta}r_0\right)
\displaystyle\left(1-\frac{3\sqrt{\theta}r_0}{2 r_h} +\frac{2\theta r^2_0}{r^2_h} \right)^2},
\end{eqnarray}
which can be rewritten as
\begin{eqnarray}
\label{tempr0}
T_H=\frac{1}{\pi r_h}\left(1+\frac{7\theta r^2_0}{r^2_h}  \right).
\end{eqnarray}
Again, we see that noncommutativity has the effect of increasing the temperature when we vary the parameter $\theta$.

In this case the entropy is now given by
\begin{eqnarray}
S=\frac{1}{4}\int \left(1- \frac{28\pi\theta r^2_0}{A}\right)dA
=\frac{1}{4}\left( A - 28\pi\theta r^2_0\ln A \right).
\end{eqnarray}
Thus, we have found a logarithmic correction for entropy.

Next to analyze the stability of the canonical acoustic black hole we will compute the specific heat
capacity using the relation~\cite{Zhang:2016pqx}
\begin{eqnarray}
C=\frac{\partial E}{\partial T_H}=\left(\frac{\partial E}{\partial r_h}\right)\left(\frac{\partial T_H}
{\partial r_h}\right)^{-1}
=T_H\left( \frac{\partial S}{\partial T_H} \right)
=-2\pi r^2_h \left(1-\frac{21\theta r^2_0}{r^2_h} \right).
\end{eqnarray}
Hence, for 
\begin{eqnarray}
r_h=\sqrt{21\theta}\,r_0
\end{eqnarray}
the specific heat capacity is zero and the noncommutative canonical acoustic black hole becomes stable with a temperature given by
\begin{eqnarray}
T_{Hmax}=\frac{4}{3\pi r_h}.
\end{eqnarray}
{In the situation where $ \theta=0 $ the specific heat
capacity is $ C=-2\pi r^2_h$ and so the canonical acoustic black hole is unstable 
by presenting a good analogy with the gravitational black hole.}
Note that due to the effect of noncommutativity via application of an external magnetic field a stability condition for the noncommutative canonical acoustic black hole is generated.

\begin{figure}[htb]
\includegraphics[scale=0.45]{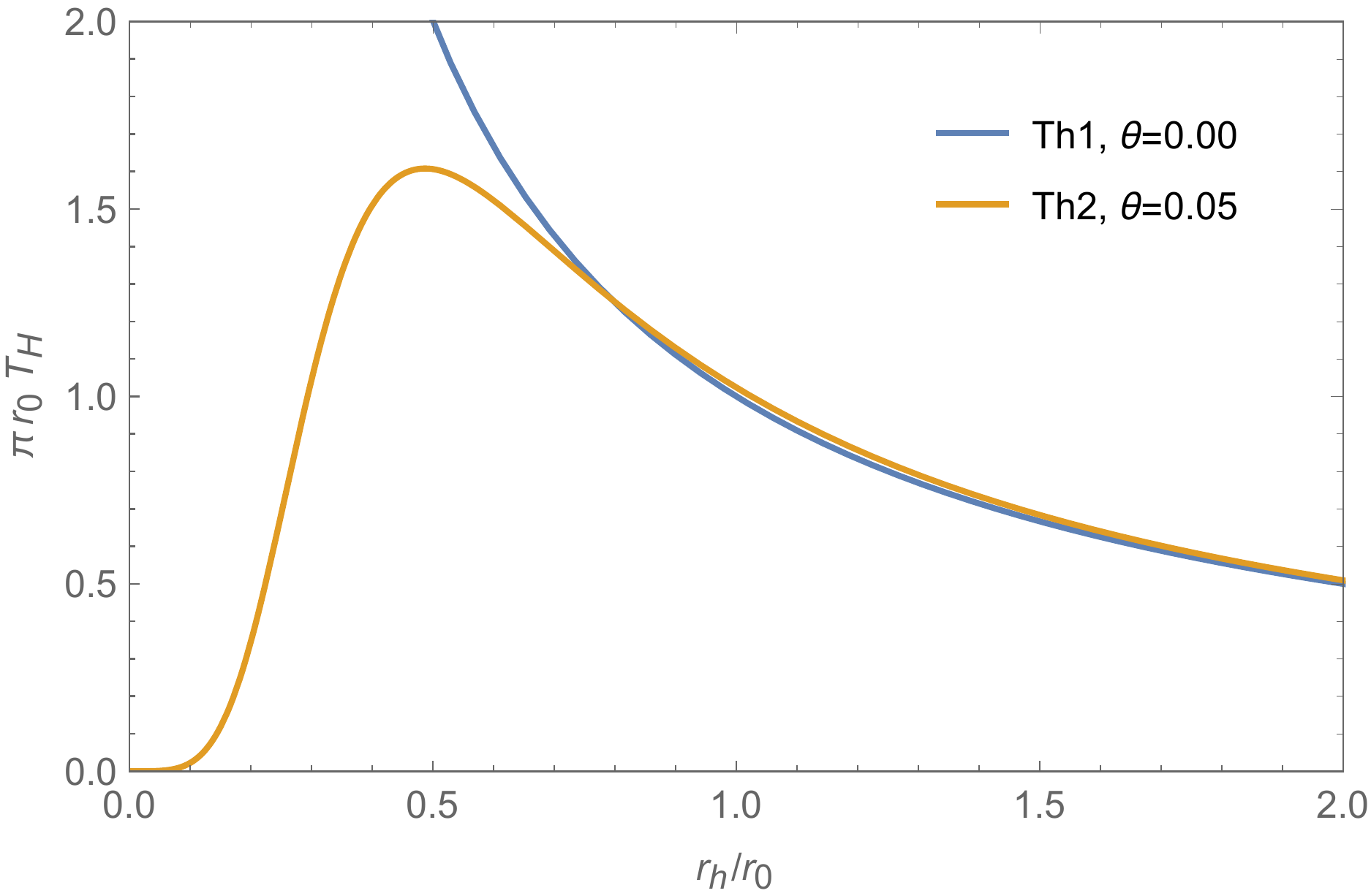}
\caption{The Hawking temperature $ \pi r_0 T_H$ in function of $ r_h/r_0 $. For $ r_h=0 $ the temperature $ Th1 $ (\ref{Th1}) diverges.
The temperature $ Th2 $ (\ref{THr0}) reaches a maximum value $ T_{max} $, and then decreases to zero as
$ r_h/r_0\rightarrow 0 $.}
\label{tht}
\end{figure}

{In the absence of noncommutativity the temperature of the canonical acoustic black hole is given by
\begin{eqnarray}
\label{Th1}
T_h=\frac{1}{\pi r_h},
\end{eqnarray}
for $ r_h=0 $ the temperature diverges.
On the other hand, the temperature of the canonical acoustic black hole (\ref{THr0}) is now regularized by the noncommutativity effect.
Thus, by taking $ r_h\rightarrow 0 $ in (\ref{THr0}) the temperature goes to zero, before it passes through a maximum point, as we can see in Fig.~\ref{tht}.}

\subsection{Electric Sector}
At this point, we will consider the situation where $B=0$ and $\theta {\cal E}_r\neq 0$.
So, from (\ref{Fr}), we have
\begin{eqnarray}
\tilde{\cal F}(r)=\left[ 1 -\theta {\cal E}_r\frac{r^2_{h}}{r^2} -\frac{r^4_{h}}{r^4}
\right]
\left[1-3\theta {\cal E}_r\frac{r^2_{h}}{r^2}\right]^{-1/2}.
\end{eqnarray}
{Here, we also have the Reissner-Nordstr\"om type correction term $\theta {\cal E}_r/r^2$ with an effective charge $Q^2=\theta {\cal E}_r$ generated due to the noncommutativity.}

We also noticed that the temperature is increased when we vary the $\theta$ parameter
\begin{equation}
\label{Theb} 
T_{H}=\frac{\left[1+\theta {\cal E}_r/2\right]}
{\sqrt{1-3\theta {\cal E}_r}}
\frac{1}{\pi r_{h}}=\frac{\left(1+2\theta {\cal E}_r  \right)}{\pi r_{h}}.
\end{equation}
For entropy we have
\begin{eqnarray}
S=\left(1-2\theta {\cal E}_r\right)\frac{A}{4}.
\end{eqnarray}
Now considering $ \theta {\cal E}_r=\theta\lambda^2/r^2 $, being $ \lambda $ a parameter with length dimension 
and $ \theta $ a dimensionless parameter, so we have
\begin{eqnarray}
\label{Fr3}
\tilde{\cal F}(r)=\left[ 1-\theta\frac{\lambda^2 r^2_h}{r^4} -\frac{r^4_{h}}{r^4} \right]
\left[1-3\theta\frac{\lambda^2 r^2_h}{r^4} \right]^{-1/2}.
\end{eqnarray}
{In this case for the Hawking temperature we find
\begin{eqnarray}
\label{Thl}
T_H=\frac{1}{\pi\displaystyle\left(r_h + 2\sqrt{\theta}\lambda +\frac{\theta \lambda^2}{2r_h}\right)
\left(1-\frac{\sqrt{\theta}\lambda}{ r_h} +\frac{\theta \lambda^2}{2r^2_h} \right)^2},
\end{eqnarray}
hence we have}
\begin{eqnarray}
T_{H}=\frac{1}{\pi r_h}\left(1+\frac{5\theta \lambda^2}{2 r^2_h}  \right).
\end{eqnarray}
{Here again, we note the effect of noncommutativity on the Hawking temperature, that is, by increasing the parameter $\theta$, the intensity of the Hawking temperature increases.}

Therefore, for entropy we obtain
\begin{eqnarray}
S=\frac{1}{4}\int \left(1- \frac{10\pi\theta \lambda^2}{ A}\right)dA
=\frac{1}{4}\left( A - 10\pi\theta\lambda^2\ln A \right).
\end{eqnarray}
Now to investigate the stability of the noncommutative canonical acoustic black hole, we determine the specific heat capacity which is given by
\begin{eqnarray}
C=-2\pi r^2_h \left(1-\frac{15\theta\lambda^2}{2r^2_h} \right).
\end{eqnarray}
The specific heat capacity is zero for 
\begin{eqnarray}
r_h=\sqrt{\frac{15\theta}{2}}\,\lambda .
\end{eqnarray}
Consequently, the noncommutative canonical acoustic black hole reaches its stability with a temperature given by
\begin{eqnarray}
T_{Hmax}=\frac{4}{3\pi r_h}.
\end{eqnarray}
Again, we verify the stability condition of the canonical acoustic black hole implemented by the effect of noncommutativity via application of an external electric field. 

\begin{figure}[htb]
\includegraphics[scale=0.45]{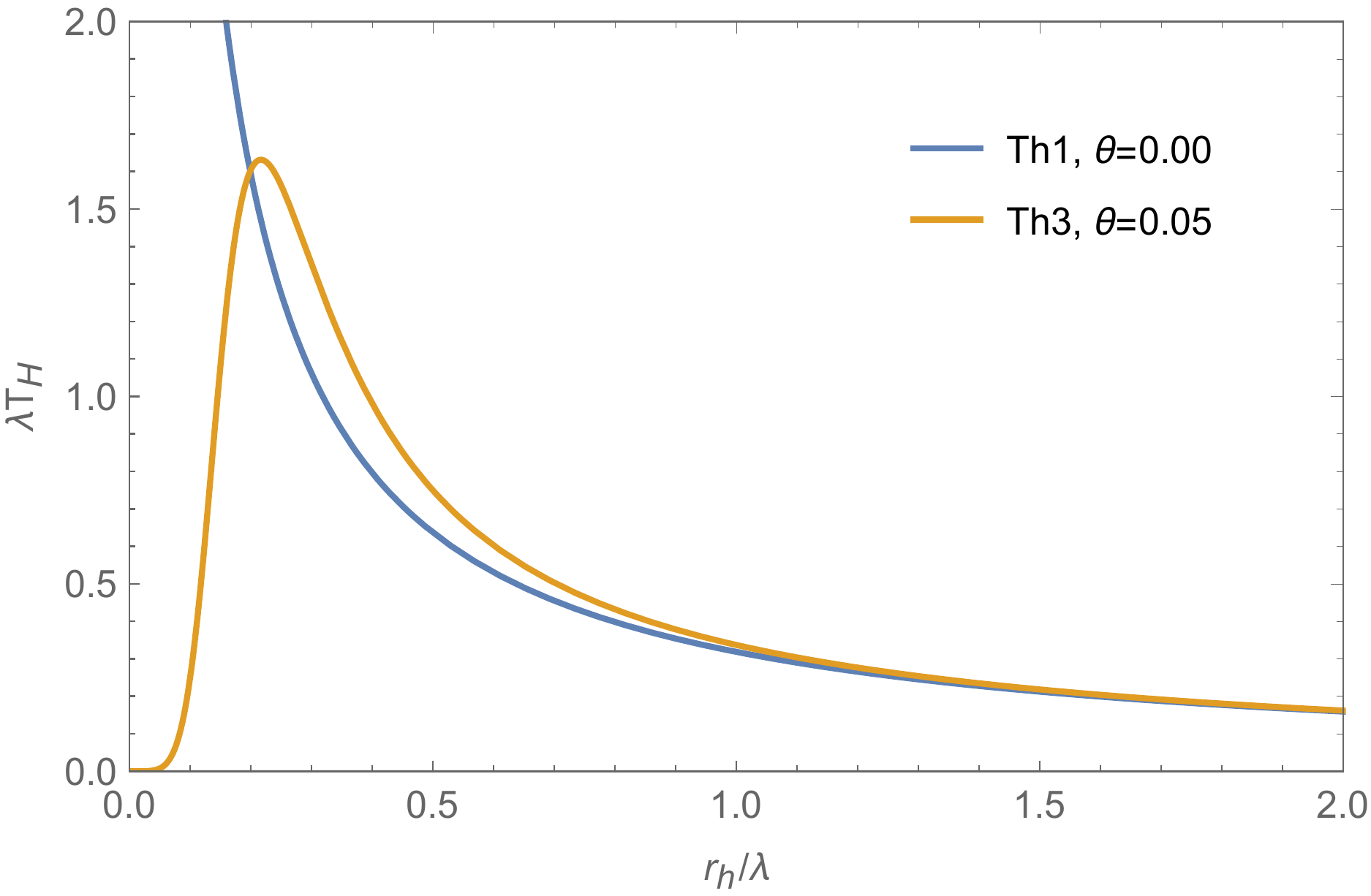}
\caption{The Hawking temperature $ \lambda T_H$ in function of $ r_h/\lambda $.
Note that the temperature $ Th3 $ (\ref{Thl}) reaches a maximum value $ T_{max} $, and then decreases to zero as
$ r_h/\lambda\rightarrow 0 $.}
\label{thl}
\end{figure}

{In Fig.~\ref{thl} we show the behavior of temperature (\ref{Thl}) as a function of $ r_h/\lambda $. 
In the limit of $ r_h/\lambda\rightarrow 0 $ the temperature passes through a maximum value and then goes to zero.}

\section{Quantum-Corrected Canonical Acoustic Black Hole}
\label{qcbh}
\begin{figure}[htb]
\includegraphics[scale=0.45]{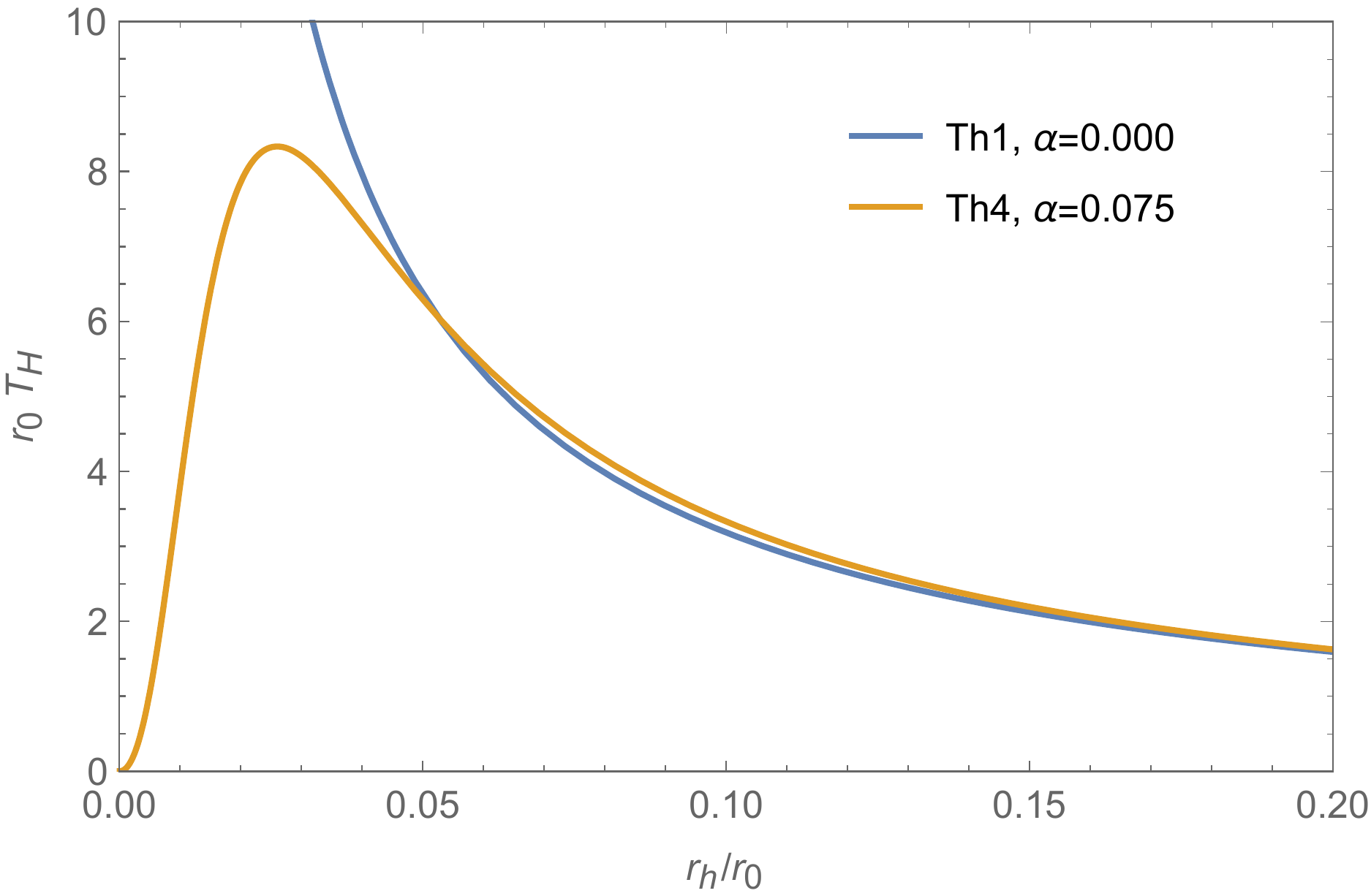}
\caption{The Hawking temperature $ r_0 T_H$ in function of $ r_h/r_0 $.
Note that the temperature $ Th4 $ (\ref{Th4}) reaches a maximum value $ T_{max} $, and then decreases to zero as
$ r_h/r_0\rightarrow 0 $.}
\label{tha}
\end{figure}

In~\cite{Anacleto:2021nhm} we have found an effective metric for acoustic black holes with quantum corrections implemented by the generalized Heisenberg uncertainty principle (GUP)~\cite{Das:2008kaa,Das:2009hs,Ali:2011fa,Ali:2009zq,Casadio:2014pia,Kempf:1994su,Nozari:2012gd,Garay:1994en,Amelino-Camelia:2000cpa,Scardigli:1999jh,Scardigli:2003kr,Scardigli:2014qka,Scardigli:2016pjs,Scardigli:2018jlm,Scardigli:2008jn,Nozari:2005ah,Nozari:2008gp,Nozari:2008rc,Nozari:2006ka} in the Abelian Higgs model.
However, the line element written in the Schwarzschild-type form is given by
\begin{eqnarray}
\label{canmeta}
ds^2=-\mathcal{F}(v_r)d\tau^2 + \frac{c^2_s\left(1+6\alpha -2\alpha e\Phi\right)}
{f(v_r)\mathcal{F}(v_r)}dr^2 
+ \frac{\left[1- 2\alpha (e\Phi -2)\right]}{\sqrt{f(v_r)}}r^2d\Omega^2,
\end{eqnarray}
where
\begin{eqnarray}
\mathcal{F}(v_r)=\frac{F(v_r)}{\sqrt{f(v_r)}}
\approx c^2_s[1-\alpha(1-e\Phi + e v_r\Lambda_r)]
-v^2_r[1-\alpha(ev_r\Lambda_r -e\Phi -1 ) ] + 2e\alpha v_r\Lambda_r.
\end{eqnarray}
The metric of the canonical acoustic black hole with quantum corrections obtained in~\cite{Anacleto:2021nhm} is given by
\begin{eqnarray}
\label{canmet}
ds^2=-\mathcal{F}(r)d\tau^2 + \frac{\left(1+6\alpha -2\alpha e\Phi\right)}
{f(r)\mathcal{F}(r)}dr^2 
+ \frac{\left[1- 2\alpha (e\Phi -2)\right]}{\sqrt{f(r)}}r^2d\Omega^2,
\end{eqnarray}
where
\begin{eqnarray}
\label{fmt}
\mathcal{F}(r)&=&
1-\alpha\left(1-e\Phi(r) + \frac{e r^2_h \Lambda_r(r)}{r^2}\right)
-\frac{ r^4_h}{r^4}\left[1-\alpha\left(\frac{e r^2_h \Lambda_r(r)}{r^2} -e\Phi(r) -1\right)\right] 
+ \frac{2e\alpha  r^2_h \Lambda_r(r)}{r^2},
\\
f(r)&=& 1+6\alpha - 2e\alpha\Phi + 2e\alpha\Lambda_r\frac{r^2_h}{r^2},
\end{eqnarray}
being $ r_h $ the radius of the event horizon, $ \alpha $  the dimensionless GUP parameter, $ \Phi $ and $ \Lambda_r $ the scalar and vector potentials respectively. 
Now choosing $\Lambda_r=0$ and $\Phi=r_0/r$ (being $ r_0$ a parameter with length dimension), 
the Hawking temperature and entropy are given respectively by
\begin{eqnarray}
\label{Th4}
T_H=\frac{1+\alpha}{\pi\displaystyle\left(r_h + 2r_h\sqrt{\frac{\alpha e r_0}{r_h}} +\frac{\alpha e r_0}{2}\right)
\left(1-\sqrt{\frac{\alpha e r_0}{2r_h}} +\frac{\alpha e r_0}{4r_h} \right)^2},
\end{eqnarray}
or
\begin{eqnarray}
\label{temp}
{T}_H=\frac{1}{\pi r_h}\left(1+\alpha + \frac{\alpha e r_0}{r_h}\right)
\qquad \Rightarrow\qquad \tilde{T}_H\approx \frac{1}{\pi r_h}\left(1 + \frac{\alpha e r_0}{r_h}\right),
\end{eqnarray}
\begin{eqnarray}
S=\int \frac{dA}{4\pi r_h \tilde{T}_H}=\frac{A}{4} - \frac{4\sqrt{\pi}\alpha e r_0\sqrt{A}}{4} + \frac{4\pi\alpha^2 e^2 r^2_0}{4}\ln\left(\frac{A}{4r^2_0}\right).
\end{eqnarray}
At this point we will analyze the stability condition of the modified canonical acoustic black hole due to quantum corrections computing the specific heat capacity. Thus, we obtain
\begin{eqnarray}
C=-2\pi r^2_h\left(1- \frac{2e\alpha r_0}{r_h}\right).
\end{eqnarray}
Note that by setting 
\begin{eqnarray}
r_h=2e\alpha r_0,
\end{eqnarray}
the specific heat capacity vanishes and the modified canonical acoustic black hole becomes stable with a temperature given by
\begin{eqnarray}
T_{Hmax}=\frac{3}{2\pi r_h}.
\end{eqnarray}
{The behavior of temperature (\ref{Th4}) as a function of $ r_h/r_0 $ can be seen in Fig.~\ref{tha}. 
In the limit of $ r_h/r_0\rightarrow 0 $ the temperature reaches a maximum value and then tends to zero.}

\begin{figure}[htb]
\includegraphics[scale=0.45]{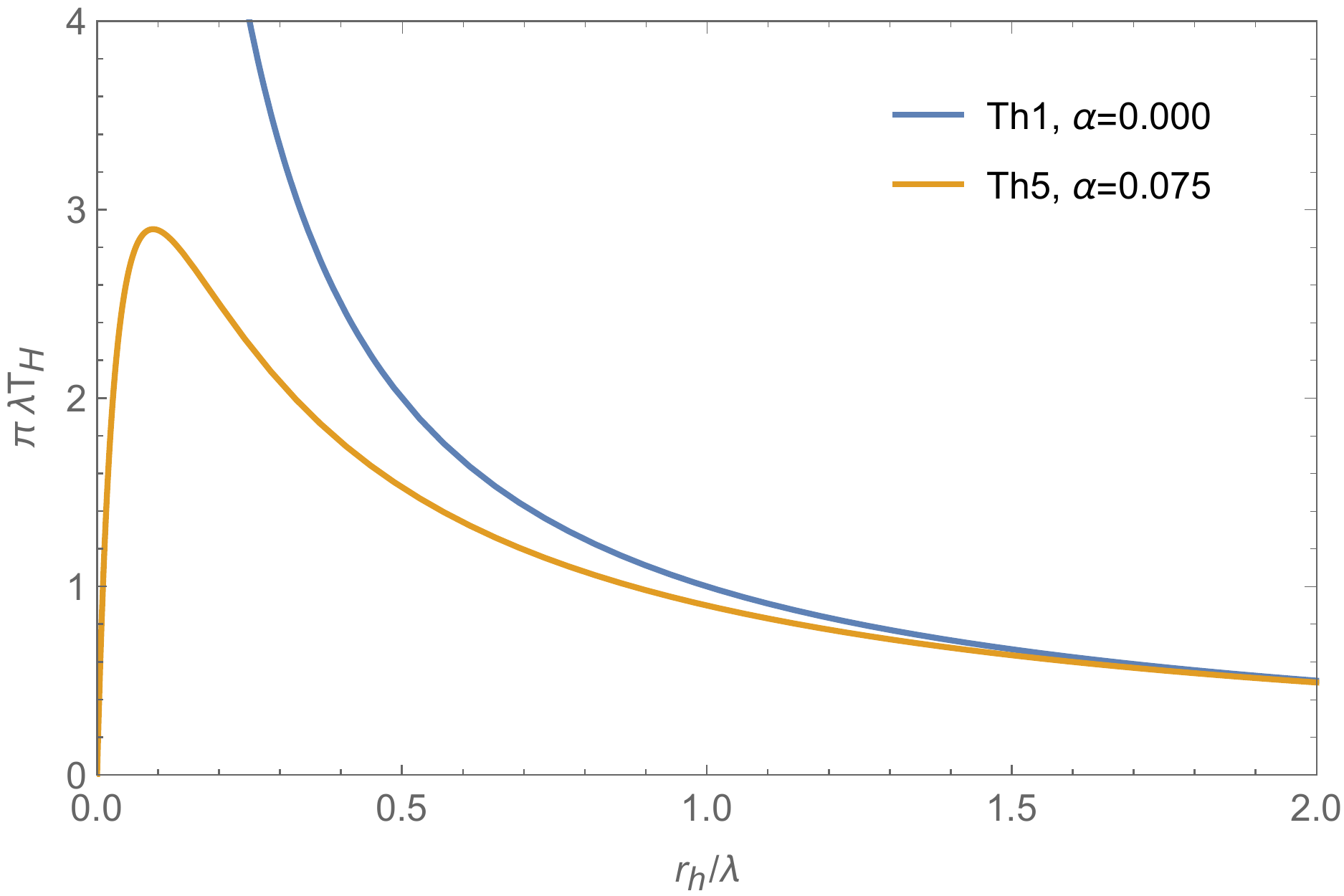}
\caption{The Hawking temperature $ \pi\lambda T_H$ in function of $ r_h/\lambda $.
Note that the temperature $ Th5 $ (\ref{Th5}) reaches a maximum value $ T_{max} $, and then decreases to zero as
$ r_h/\lambda\rightarrow 0 $.}
\label{thav}
\end{figure}

Next, by choosing $ \Phi=0 $ and $ \Lambda_r=\lambda/r $ (being $ \lambda $  parameter with length dimension), 
{we have~\cite{Anacleto:2021nhm} }
\begin{eqnarray}
\label{Th5}
T_H=\frac{1+\alpha}{\pi\displaystyle\left(r_h +\frac{3\alpha e \lambda}{2}\right)
\left(1+\frac{\alpha e \lambda}{r_h} \right)},
\end{eqnarray}
or
\begin{eqnarray}
\label{temp2}
{T}_H=T_H\left(1+\alpha - \frac{5\alpha e\lambda}{2r_h}\right)
\qquad\Rightarrow\qquad \tilde{T}_H=\frac{1}{\pi r_h}\left(1- \frac{5\alpha e\lambda}{2r_h}\right),
\end{eqnarray}
and
\begin{eqnarray}
S=\frac{A}{4} + \frac{10\sqrt{\pi}\alpha e \lambda\sqrt{A}}{4} + \frac{25\pi\alpha^2 e^2 \lambda^2}{4}\ln\left(\frac{A}{4\lambda^2}\right).
\end{eqnarray}
{For the specific heat capacity we find
\begin{eqnarray}
C=-2\pi r^2_h\left(1-\sqrt{\frac{5e\alpha\lambda}{r_h}}\right)\left(1+ \sqrt{\frac{5e\alpha\lambda}{r_h}}\right).
\end{eqnarray}
In this situation, for 
\begin{eqnarray}
\label{scl}
r_h=5e\alpha\lambda,
\end{eqnarray}
the specific heat capacity vanishes. Thus, indicating that the modified canonical acoustic black hole become stable. 
Then, replacing (\ref{scl}) in (\ref{temp2}), we find the maximum temperature given by
\begin{eqnarray}
T_{Hmax}=\frac{1}{2\pi r_h}.
\end{eqnarray}
Therefore, due to the GUP effect, we obtain the stability conditions of the canonical acoustic black hole via applications of the scalar and vector potentials. 
Note that the temperature (\ref{Th5}) is regularized by the effect of the GUP as we can see in Fig.~\ref{thav}.
For $ r_h/\lambda\rightarrow 0 $ the temperature reaches a maximum value before going to zero.}

{It is worth mentioning that the stability condition analyzed here for the noncommutative and GUP cases shows similarity with the result that arises in the gravitational case. Black hole remnant formation emerges from the minimum radius (or minimum mass) condition that nullifies the specific heat capacity, and thus, the black hole stops to evaporate completely~\cite{Nozari:2005ah}. 
Therefore, in our results we would have, for a certain minimum radius, an analogous condition for the formation of an acoustic black hole remnant due to the effect of noncommutativity and GUP.}
{In addition, we have found logarithmic corrections for entropy in the noncommutative and GUP cases. We highlight that the logarithmic correction term for entropy is fundamental to exploring the process of evaporation and formation of gravitational black hole remnant and, analogously,  we have verified this possibility in the physics of a canonical acoustic black hole in the 
noncommutative background and with GUP.}

{Also, as shown in~\cite{Anacleto:2021nhm}, we point out that by making $ \Phi=r_0/r $ and $ \Lambda_r=\lambda_0 $ (a dimensionless constant) in (\ref{fmt}), we find}
\begin{eqnarray}
\mathcal{F}(r)=
1-\alpha\left(1-\frac{er_0}{r} - \frac{e\lambda_0 r^2_h}{r^2}\right)
-\frac{ r^4_h}{r^4}\left[1 + \alpha\left(1 + \frac{er_0}{r}- \frac{e\lambda_0 r^2_h}{r^2}\right)\right],
\end{eqnarray}
{with contributions from a Reissner-Nordstr\"om type metric being generated for the canonical acoustic black hole metric function. 
Hence we identify $ Q^2= \alpha e\lambda_0 r^2_h$ as the effective charge. 
}

\section{conclusions}
\label{conclu}

In summary, by considering the noncommutative canonical acoustic black hole metric we found the Hawking temperature, 
entropy and we have verified the stability condition by calculating the specific heat capacity. 
We have been analyzing the stability in the electric and magnetic sectors. In both sectors we show that stability occurs for a certain value of the minimum horizon radius.
In the case of the canonical acoustic black hole with quantum corrections we have determined the Hawking temperature and entropy. 
In the situation where the metric depends only on the correction due to the scalar potential, we have shown that for a given value of the horizon radius the specific heat capacity goes to zero, and the canonical acoustic black hole becomes stable.
{Besides, when the metric depends only on the correction due to the vector potential, we have also found that the specific heat capacity vanishes, and the canonical acoustic black hole becomes stable. 
In addition, for the cases where stability was analyzed, we have shown that temperatures reach maximum values before going to zero when the horizon radius is zero.}

\acknowledgments

We would like to thank CNPq, CAPES and CNPq/PRONEX/FAPESQ-PB (Grant nos. 165/2018 and 015/2019),
for partial financial support. MAA, FAB and EP acknowledge support from CNPq (Grant nos. 306398/2021-4, 309092/2022-1, 304290/2020-3).

\end{document}